\documentclass[12pt,english]{article}
\usepackage[utf8x]{inputenc}
\usepackage{amssymb}
\usepackage{amsfonts}
\usepackage{amsmath}
\usepackage{graphicx}

\usepackage[svgnames]{xcolor}
\usepackage[colorlinks=true,linkcolor=Blue,citecolor=Blue,urlcolor=Blue]{hyperref}

\usepackage{babel}

\title{Slowly rotating Curzon-Chazy Metric}

\author{Paulo Montero-Camacho\thanks{School of Physics, 
University of Costa Rica, email: paulo.montero@ucr.ac.cr} \\ 
Francisco Frutos-Alfaro \\ 
Carlos Guti\'errez-Chaves} 

\date{\today}

\begin{document}

\maketitle

\begin{abstract}
A new rotation version of the Curzon-Chazy metric is found. This new metric 
was obtained by means of a perturbation method, in order to include slow 
rotation. The solution is then proved to fulfill the Einstein field equations 
using a REDUCE program. Furthermore, the applications of this new solution are 
discussed. 
\end{abstract}


\section{Introduction}

\noindent
The Curzon-Chazy metric \cite{Chazy,Curzon,Synge} is one of the simplest 
solutions of the Einstein field equations (EFE) for the Weyl metric. 
The original idea of Curzon \cite{Curzon} and Chazy \cite{Chazy} was the 
superposition of two particles at different points on the symmetry axis. 
This superposition exhibit a singularity between these particles along 
this symmetry axis. This singularity is interpreted as a strut (Weyl strut), 
which stress holds these particles apart and does not exert a gravitational 
field \cite{Belinski,Griffiths}.
  
\noindent
For one particle, the Curzon-Chazy metric describes the exterior field of 
a finite source \cite{Griffiths}, and has a spherically symmetric Newtonian 
potential of a point particle located at $ r = 0 $. The resulting spacetime 
is not spherically symmetric and its weak limit is that of an object located 
at the origin with multipoles. Moreover, the singularity at $ r = 0 $ has 
a very interesting but complicated directionally dependent structure 
\cite{Gautreau,Griffiths}. There is a curvature singularity at 
$ \rho = 0, \, z = 0 $ that is not surrounded by a horizon, 
{\it i.e.} it is naked. Every light ray emitted from it 
becomes infinitely redshifted, so that it is effectively invisible 
\cite{Griffiths}. Studying the principal null directions, it was found that 
this spacetime has an invariantly hypersurface $ \sqrt{\rho^2 + z^2} = M $, 
on which the Weyl invariant $ {\cal J} $ (determinant of the Weyl five 
complex scalar functions) vanishes \cite{Arianrhod,Griffiths}. Furthermore, 
this metric is Petrov type $ D $, except at two points ($ z = \pm M $) that 
intersect the axis $ \rho = 0 $, where it is of Petrov type $ O $ 
\cite{Arianrhod}.  

\noindent
The properties of this solution have been analyzed since its discovery such as 
the potential surfaces for its time-like geodesics and their variations with 
the change in energy \cite{Felice1990}. A generalization of the metric to 
the Einstein-Maxwell equations have also been obtained \cite{Hernandez1993}. 
A modified Curzon-Chazy metric, using a rotating reference frame as approach, 
has already been applied to study binary pulsar systems \cite{Wanas2012}.

\noindent
A slowly rotating version of the Curzon-Chazy solution could be used, 
for instance, to study the gravitational lens effect, because of its 
asymmetrical nature. Furthermore, it is important to mention Halilsoy's 
research \cite{Halilsoy}, in his work he obtained a rotating Curzon metric 
using the Ernst potential method \cite{Ernst1968}, however the rotational 
metric component term ($ g_{03} $) obtained in his work is non flat, therefore 
it could not be of astrophysical interest. Besides there is a problem with the 
generalization Halilsoy  proposes for an arbitrary number of massive particules 
it fails to obtain the correct form for the case of two massive particles. 

\noindent
In this work, a slow rotating metric is obtained by introducing a perturbation 
in the metric rather than using a rotating reference frame. 
Our rotational metric term is Kerr like. Moreover, we discuss the possible 
applications of this new version.

\section{The Curzon-Chazy Metric}

\noindent
The Curzon-Chazy metric in canonical cylindrical coordinates is given by
\cite{Carmeli,Chazy,Curzon} (in geometrical units $ G=c=1 $):

\begin{equation}
\label{curzon1}
d s^2 = e^{- 2 \psi} d t^2 - e^{2 (\psi - \gamma)} (d {\rho}^2 + d z^2) 
- e^{2 \psi} {\rho}^2 d {\phi}^2 
\end{equation}

\noindent
where

\begin{eqnarray}
\label{expres1}
\psi   & = & \frac{M}{\eta} , \nonumber \\
\gamma & = & \frac{M^2 {\rho}^2}{2 \eta^4} , \nonumber \\
\eta^2 & = & {\rho}^2 + z^2 . \nonumber \\
\end{eqnarray}

\noindent
The Curzon-Chazy metric in spherical coordinates can be obtained by means of 
the following mapping \cite{Carmeli}:

\begin{equation} 
\label{mapping}
\rho = \sqrt{Z} \quad {\rm and} \quad z = (r - M) \cos \theta , 
\end{equation}

\noindent
where $ Z = (r^2 - 2Mr) \sin^2{\theta} $ 

\noindent
Using this transformation the Curzon-Chazy metric takes the form

\begin{equation}
\label{curzon2}
d s^2 = e^{-2 \psi} d t^2 - e^{2(\psi - \gamma)} (X d r^2 + Y d {\theta}^2) 
- e^{2 \psi} Z d {\phi}^2 
\end{equation}

\noindent
where

\begin{eqnarray}
\label{expres2}
\psi   & = & \frac{M}{\eta} , \nonumber \\
\gamma & = & \frac{M^2 (r^2 -2Mr) \sin^2 \theta}{2 \eta^4} , \nonumber \\
\eta^2   & = & r^2 - 2Mr + M^2 \cos^2{\theta} \nonumber \\
\Delta & = & r^2 - 2Mr + M^2 \sin^2{\theta} , \\
X      & = & \frac{\Delta}{r^2 - 2Mr}, \nonumber \\
Y      & = & \Delta  \nonumber 
\end{eqnarray}

\section{The Lewis Metric}

\noindent
The Lewis metric is given by \cite{Lewis,Carmeli} 

\begin{equation}
\label{lewis}
d s^2 = V d t^2 - 2W dt d \phi -  e^{\mu} d {\rho}^2 -  e^{\nu} dz^2 
- \Sigma d {\phi}^2
\end{equation}

\noindent
where we have chosen the canonical coordinates $ x^1 = \rho $ and $ x^2 = z $. 
The potentials $ V, \, W, \, \Sigma, \, \mu $ and $ \nu $ are functions of 
$ \rho $ and $ z \, (\rho^2 = V \Sigma + W^2) $. Choosing $ \mu = \nu $ and 
performing the following changes of potentials

\begin{align}
V & = f, &  W & = \omega f, &  \Sigma & = \frac{\rho^2}{f} - \omega^2 f, 
&   e^\mu & = \frac{e^\chi}{f} &
\end{align}

\noindent
we get the Papapetrou metric

\begin{equation}
\label{papapetrou}
d s^2 = f (dt - \omega d \phi)^2 - \frac{e^\chi}{f} [d {\rho}^2 + d z^2] 
- \frac{\rho^2}{f} d {\phi}^2 
\end{equation}

\noindent
Note that for slow rotation we neglect the second order in $\omega $, hence \\
$ \omega^2 \simeq 0 \Rightarrow W^2 \simeq 0 $, and 
$ \Sigma \simeq \rho^2 / f $.

\section{Perturbing the Curzon-Chazy Metric}

\noindent
To include slow rotation into the Curzon-Chazy metric we choose the 
Lewis-Papapetrou metric, equation (\ref{papapetrou}). First of all, we choose 
expressions for the canonical coordinates $ \rho $ and $ z $. 
From (\ref{mapping}) we get 

\begin{equation}
\label{rhotheta}
d {\rho}^2 + d z^2 = \Delta \left(\frac{d r^2}{r^2 - 2Mr} 
+ d {\theta}^2 \right) 
= X dr^2 + Y d {\theta}^2 .
\end{equation}  

\noindent 
Now, let us choose $ V = f = {\rm e}^{-2 \psi} $ and neglect the second order in 
$ \omega $. Then, we have

$$ \Sigma \simeq \frac{\rho^2}{f} = Z e^{2\psi} \quad {\rm and} \quad 
\frac{e^\chi}{f} = e^{2(\psi - \gamma)} . $$

\noindent 
The metric takes the form

\begin{equation}
\label{metric1}
d s^2 = e^{-2 \psi} d t^2 - 2 W dt d \phi 
- e^{2(\psi - \gamma)}[X dr^2 + Y d {\theta}^2] - Z e^{2\psi} d {\phi}^2
\end{equation}

\noindent 
To obtain a slowly rotating version of the metric (\ref{curzon2}), the only 
potential, we have to find is $ W $. In order to do this, we need to solve the 
Einstein equations for this metric

\begin{equation}
G_{ij} = R_{ij} - \frac{R}{2} g_{ij} = 0
\end{equation}

\noindent 
where $ R_{ij} \, (i, \, j = 0, \, 1, \, 2, \, 3) $ are the Ricci tensor 
components and $ R $ is the curvature scalar. To find the approximated slowly 
version of the metric, we wrote a REDUCE program to find the Ricci tensor.
The interested reader can request this program by sending us an message. 
Fortunately, the Ricci tensor components $ R_{00}, R_{11}, R_{12}, R_{22}, $ 
$ R_{23}, R_{33} $ and the scalar curvature depend upon the potentials 
$ V, \, X, \, Y, \, Z $ and not on $ W $ (see Appendix). Hence, these 
components vanish. The only equation we have to solve is $ R_{03} = 0 $, 
because it depends upon $ W $. The equation for this Ricci component, up to 
order $ O(M^3, \, a^2) $, is

\begin{equation}
\label{eqdif}
\sin{\theta} \left(\frac{\partial^2 W}{\partial \theta^2} 
+ r^2 \frac{\partial^2 W}{\partial r^2} \right) 
- \cos{\theta} \frac{\partial W}{\partial \theta} = 0 .
\end{equation}

\noindent 
The solution for (\ref{eqdif}) is

\begin{equation}
W = \frac{K}{r} \sin^2 \theta
\end{equation}

\noindent 
In order to find the constant $ K $  let us see the Lense-Thirring metric 
which is obtained from the Kerr metric: 

\begin{equation}
\label{lense}
d s^2 = \left(1 - \frac{2M}{r} \right) d t^2 
+ \frac{4 J}{r} \sin^2{\theta}  d t d \phi 
- \left(1 - \frac{2M}{r} \right)^{-1} d r^2 - r^2 d {\Sigma}^2
\end{equation}

\noindent 
where $ d \Sigma^2 = d {\theta}^2 + \sin^2{\theta} \, d {\phi}^2 $ and 
$ J = M a $ is the angular momentum. At first order in $ M $, this metric and 
the Curzon-Chazy metric coincides. Then, comparing the second term of the 
latter metric with the corresponding term of (\ref{lense}), we note that 
$ K = - 2 J = - 2 M a $.

\noindent 
The new rotating Curzon-Chazy metric is

\begin{eqnarray}
\label{metric2}
d s^2 & = & e^{-2 \psi} d t^2 + \frac{4 J}{r} \sin^2{\theta}  d t d \phi 
- \Delta e^{2(\psi - \gamma)} 
\left(\frac{d r^2}{r^2 - 2Mr} + d {\theta}^2 \right) \\ 
& - & e^{2 \psi} (r^2 - 2Mr) \sin^2{\theta} d {\phi}^2 . \nonumber  
\end{eqnarray}

\noindent 
We check that the metric (\ref{metric2}) is indeed a solution of the Einstein's 
Field Equations, up to the order $ O(M^3, \, a^2) $, using the same REDUCE 
program.

\section{Discussion and applications of the Metric}

\noindent
The slow rotating solution here presented, needs to be analyzed in 
the same manner as the static Curzon-Chazy metric has been.

\noindent
The new rotating metric can be used to describe two bodies rotating around its 
center of mass, performing orbits. This system can then be applied to 
significant binary star systems. 

\noindent
New calculations, such as the geodesics can be performed in order to visualize 
the trajectories due to such gravitational field. Also, a description of the 
surface potentials due to these geodesics can be studied, as it was done for 
the non-rotating Curzon-Chazy solution \cite{Felice1990}.

\noindent 
Moreover, in treating with two singularities, gravitational lens calculations 
can be performed upon this system; equally the Lense-Thirring correction could 
certainly be applied, which results will lead to a more real scenario.

\noindent 
For Postnewtonian calculations, for example in gravitational lens theory and 
astrometry, we need to expand the metric (\ref{metric2}) in a Taylor series, 
the result is

\begin{eqnarray}
\label{ppn}
{d} {s}^2 & = & 
\left(1 - \frac{2 M}{r} 
+ \frac{2}{3} \frac{M^3}{r^3} P_2 (\cos{\theta}) \right) {d} {t}^2 
+ \frac{4 J}{r} \sin^2{\theta}  d t d \phi \nonumber \\
& - & \left(1 + \frac{2 M}{r} + \frac{4 M^2}{r^2} 
+ 2 \left(4 - \frac{1}{3} P_2(\cos{\theta})\right) \frac{M^3}{r^3} \right) 
{d} {r}^2 \\
& - & r^2 \left(1 - \frac{2}{3} \frac{M^3}{r^3} P_2(\cos{\theta}) \right) 
{d} \Sigma^2 , \nonumber 
\end{eqnarray}

\noindent
where $ P_2(\cos{\theta}) = {(3 \cos^2{\theta} - 1)}/{2} $ and 
$ {d} \Sigma^2 = {d} {\theta}^2 + \sin^2{\theta} {d} {\phi}^2 $.
It was kept the series till the third order in $ M $ to take into account 
the quadrupole moment.

\noindent
For other purposes, for instance Astrometry, it is usual to transform this 
metric (\ref{ppn}) into cartesian coordinates using the harmonic or the 
isotropic coordinates of Schwarzschild metric. The first one is 
$ r = {\bar{r}} + M $, and the second one is 
$ r = {\bar{r}}(1 + {M}/{2 {\bar{r}}})^2 $, 
where $ {\bar{r}} $ is a new radial coordinate \cite{Weinberg}. 
Using the first one, the metric (\ref{ppn}) becomes (dropping the bars)

\begin{eqnarray}
\label{ppn2}
{d} {s}^2 & = & 
\left(1 - \frac{2 M}{r} + \frac{2 M^2}{r^2}
+ 2 \left(\frac{1}{3} P_2 (\cos{\theta}) - 1 \right) 
\frac{M^3}{r^3} \right) {d} {t}^2 
+ \frac{4 J}{r} \sin^2{\theta}  d t d \phi \nonumber \\
& - & \left(1 + \frac{2 M}{r} +  \frac{2 M^2}{r^2} 
+ 2 \left(1 - \frac{1}{3} P_2(\cos{\theta}) \right) \frac{M^3}{r^3} \right) 
{d} {r}^2 \\
& - & r^2 \left(1 + \frac{M}{r} \right)^2 
\left(1 - \frac{2}{3} \frac{M^3}{r^3} P_2(\cos{\theta}) \right) {d} \Sigma^2 . 
\nonumber 
\end{eqnarray}

\noindent
Now, the transformation into cartesian coordinates gives the following result:

\begin{eqnarray}
\label{ppn3}
{d} {s}^2 & = & 
\left(1 - \frac{2 M}{r} + \frac{2 M^2}{r^2}
+ 2 \left(\frac{1}{3} P_2 (\cos{\theta}) - 1 \right) 
\frac{M^3}{r^3} \right) {d} {t}^2 \nonumber \\
& + &  \frac{4 J}{r} \sin^2{\theta}  d t (x d y - y d x)  \\ 
& - & \left(1 + \frac{2 M}{r} +  \frac{M^2}{r^2} 
- \frac{M^3}{3 r^3} P_2(\cos{\theta}) \right) {d} {\bf {x}}^2 \nonumber \\
& - & \frac{M^2}{r^2} \left(1 + \frac{2 M}{r} \right) 
[{\bf {x}} \cdot {d}{\bf {x}} ]^2 , \nonumber 
\end{eqnarray}
 
\noindent
where

\begin{eqnarray}
x & = & r \sin{\theta} \cos{\phi} \\ \nonumber 
y & = & r \sin{\theta} \sin{\phi} \\ \nonumber 
z & = & r \cos{\theta} \nonumber 
\end{eqnarray}

\noindent
The metric can be generalized for any $ J $ direction using the vector 
$ {\bf V} $ defined by

$$ {\bf V} = \frac{G}{2 c^3 r^3} [{\bf {J}} \times {\bf {x}}] , $$

\noindent
with $ {\bf {J}} = J {\bf {e}}_j $ 
($ {\bf {e}}_j $ is an unit vector in the direction of $ {\bf {J}} $).
Then, the metric (\ref{ppn3}) takes the form

\begin{eqnarray}
\label{ppn4}
{d} {s}^2 & = & 
\left(1 - \frac{2 M}{r} + \frac{2 M^2}{r^2}
+ 2 \left(\frac{1}{3} P_2 (\cos{\theta}) - 1 \right) 
\frac{M^3}{r^3} \right) {d} {t}^2 
+ 8 {\bf {V}} \cdot d {\bf {x}} d t \nonumber \\
& - & \left(1 + \frac{2 M}{r} +  \frac{M^2}{r^2} 
- \frac{M^3}{3 r^3} P_2(\cos{\theta}) \right) {d} {\bf {x}}^2 \\
& - & \frac{M^2}{r^2} \left(1 + \frac{2 M}{r} \right) 
[{\bf {x}} \cdot {d}{\bf {x}} ]^2 . \nonumber 
\end{eqnarray}
 
\noindent
The metric (\ref{ppn4}) is ready for calculations on astrometry and 
gravitational lensing.
  

\newpage

\appendix
\section{Appendix}

\noindent 
The Ricci tensor components are

\begin{eqnarray*}
R_{0 0} & = & \frac{{\rm e}^{2 (\gamma - 2 \psi)}}{2 X^2 Y^2 Z} \left( 
- 2 X^2 Y Z \frac{\partial^2 \psi}{\partial \theta^2} 
- X Y Z \frac{\partial \psi}{\partial \theta} 
\frac{\partial X}{\partial \theta} 
+ X^2 Z \frac{\partial \psi}{\partial \theta} 
\frac{\partial Y}{\partial \theta} \right. \\
& - & \left. 
X^2 Y \frac{\partial \psi}{\partial \theta} \frac{\partial Z}{\partial \theta} 
- 2 X Y^2 Z \frac{\partial^2 \psi}{\partial r^2} 
+ Y^2 Z \frac{\partial \psi}{\partial r} \frac{\partial X}{\partial r} 
-  X Y Z \frac{\partial \psi}{\partial r} \frac{\partial Y}{\partial r} 
\right. \\
& - & \left. X Y^2\frac{\partial \psi}{\partial r} 
\frac{\partial Z}{\partial r} \right) \\
R_{0 1} & = & 0 \\
R_{0 2} & = & 0 \\
R_{0 3} & = & \frac{{\rm e}^{2 (\gamma - \psi)}}{2 X^2 Y^2 Z} \left( 
- X Y Z \frac{\partial X}{\partial \theta} \frac{\partial W}{\partial \theta} 
+ Y^2 Z \frac{\partial X}{\partial r} \frac{\partial W}{\partial r} 
- 2 X^2 Y Z \frac{\partial^2 W}{\partial \theta^2} \right. \\
& + & \left. X^2 Z \frac{\partial W}{\partial \theta} 
\frac{\partial Y}{\partial \theta} 
+ X^2 Y \frac{\partial W}{\partial \theta} \frac{\partial Z}{\partial \theta} 
- 2 X Y^2 Z \frac{\partial^2 W}{\partial r^2} 
- X Y Z \frac{\partial W}{\partial r} \frac{\partial Y}{\partial r} \right. \\
& + & \left. X Y^2 \frac{\partial W}{\partial r} 
\frac{\partial Z}{\partial r} \right) \\
R_{1 1} & = & \frac{1}{4 X Y^2 Z^2}\left(
- 4 X^2 Y Z^2 \frac{\partial^2 \psi}{\partial \theta^2} 
- 2 X Y Z^2 \frac{\partial \psi}{\partial \theta} 
\frac{\partial X}{\partial \theta} 
+ 2 X^2 Z^2 \frac{\partial \psi}{\partial \theta} 
\frac{\partial Y}{\partial \theta} \right. \\
& - & \left. 2 X^2 Y Z \frac{\partial \psi}{\partial \theta} 
\frac{\partial Z}{\partial \theta} 
- 4 X Y^2 Z^2 \frac{\partial^2 \psi}{\partial r^2} 
- 8 X Y^2 Z^2 \left[\frac{\partial \psi}{\partial r}\right]^2 
+ 2 Y^2 Z^2 \frac{\partial \psi}{\partial r} 
\frac{\partial X}{\partial r} \right. \\
& - & \left. 2 X Y Z^2 \frac{\partial \psi}{\partial r} 
\frac{\partial Y}{\partial r} 
- 2 X Y^2 Z \frac{\partial \psi}{\partial r} \frac{\partial Z}{\partial r} 
+ 4 X^2 Y Z^2 \frac{\partial^2 \gamma}{\partial \theta^2} 
+ 2 X Y Z^2 \frac{\partial \gamma}{\partial \theta} 
\frac{\partial X}{\partial \theta} \right. \\ 
& - & \left. 2 X^2 Z^2 \frac{\partial \gamma}{\partial \theta} 
\frac{\partial Y}{\partial \theta} 
+ 2 X^2 Y Z \frac{\partial \gamma}{\partial \theta} 
\frac{\partial Z}{\partial \theta} 
+ 4 X Y^2 Z^2 \frac{\partial^2 \gamma}{\partial r^2} 
- 2 Y^2 Z^2 \frac{\partial \gamma}{\partial r} 
\frac{\partial X}{\partial r} \right. \\
& + & \left. 2 X Y Z^2 \frac{\partial \gamma}{\partial r} 
\frac{\partial Y}{\partial r} 
- 2 X Y^2 Z \frac{\partial \gamma}{\partial r} \frac{\partial Z}{\partial r} 
- 2 X Y Z^2 \frac{\partial^2 X}{\partial \theta^2} 
+ Y Z^2 \left[\frac{\partial X}{\partial \theta}\right]^2 \right. \\
& + & \left. X Z^2 \frac{\partial X}{\partial \theta} 
\frac{\partial Y}{\partial \theta} 
- X Y Z \frac{\partial X}{\partial \theta} \frac{\partial Z}{\partial \theta} 
+ Y Z^2 \frac{\partial X}{\partial r} \frac{\partial Y}{\partial r} 
+ Y^2 Z \frac{\partial X}{\partial r} \frac{\partial Z}{\partial r} \right. \\
& - & \left. 2 X Y Z^2 \frac{\partial^2 Y}{\partial r^2} 
+ X Z^2 \left[\frac{\partial Y}{\partial r}\right]^2 
- 2 X Y^2 Z \frac{\partial^2 Z}{\partial r^2} 
+ X Y^2 \left[\frac{\partial Z}{\partial r}\right]^2 \right) 
\end{eqnarray*}

\begin{eqnarray*}
R_{1 2} & = & \frac{1}{4 X Y Z^2} \left(
- 8 X Y Z^2 \frac{\partial \psi}{\partial \theta} 
\frac{\partial \psi}{\partial r} 
- 2 X Y Z \frac{\partial \gamma}{\partial \theta} \frac{\partial Z}{\partial r} 
- 2 X Y Z \frac{\partial \gamma}{\partial r} 
\frac{\partial Z}{\partial \theta} \right. \\
& + & \left. Y Z \frac{\partial X}{\partial \theta} 
\frac{\partial Z}{\partial r} 
+ X Z \frac{\partial Y}{\partial r} \frac{\partial Z}{\partial \theta} 
- 2 X Y Z \frac{\partial^2 Z}{\partial \theta \partial r} 
+ X Y \frac{\partial Z}{\partial \theta} 
\frac{\partial Z}{\partial r} \right) \\
R_{1 3} & = & 0 \\
R_{2 2} & = & \frac{1}{4 X^2 Y Z^2} \left(
- 4 X^2 Y Z^2 \frac{\partial^2 \psi}{\partial \theta^2} 
- 8 X^2 Y Z^2 \left[\frac{\partial \psi}{\partial \theta}\right]^2 
- 2 X Y Z^2 \frac{\partial \psi}{\partial \theta} 
\frac{\partial X}{\partial \theta} \right. \\
& + & \left. 2 X^2 Z^2 \frac{\partial \psi}{\partial \theta} 
\frac{\partial Y}{\partial \theta} 
- 2 X^2 Y Z \frac{\partial \psi}{\partial \theta} 
\frac{\partial Z}{\partial \theta} 
- 4 X Y^2 Z^2 \frac{\partial^2 \psi}{\partial r^2} 
+ 2 Y^2 Z^2 \frac{\partial \psi}{\partial r} 
\frac{\partial X}{\partial r} \right. \\
& - & \left. 2 X Y Z^2 \frac{\partial \psi}{\partial r} 
\frac{\partial Y}{\partial r} 
- 2 X Y^2 Z \frac{\partial \psi}{\partial r} \frac{\partial Z}{\partial r} 
+ 4 X^2 Y Z^2 \frac{\partial^2 \gamma}{\partial \theta^2} 
+ 2 X Y Z^2 \frac{\partial \gamma}{\partial \theta} 
\frac{\partial X}{\partial \theta} \right. \\
& - & \left. 2 X^2 Z^2 \frac{\partial \gamma}{\partial \theta} 
\frac{\partial Y}{\partial \theta} 
- 2 X^2 Y Z \frac{\partial \gamma}{\partial \theta} 
\frac{\partial Z}{\partial \theta} 
+ 4 X Y^2 Z^2 \frac{\partial^2 \gamma}{\partial r^2} 
- 2 Y^2 Z^2 \frac{\partial \gamma}{\partial r} 
\frac{\partial X}{\partial r} \right. \\
& + & \left. 2 X Y Z^2 \frac{\partial \gamma}{\partial r} 
\frac{\partial Y}{\partial r} 
+ 2 X Y^2 Z \frac{\partial \gamma}{\partial r} \frac{\partial Z}{\partial r} 
- 2 X Y Z^2 \frac{\partial^2 X}{\partial \theta^2} 
+ Y Z^2 \left[\frac{\partial X}{\partial \theta}\right]^2 \right. \\ 
& + & \left. X Z^2 \frac{\partial X}{\partial \theta} 
\frac{\partial Y}{\partial \theta} 
+ Y Z^2 \frac{\partial X}{\partial r} \frac{\partial Y}{\partial r} 
+ X^2 Z \frac{\partial Y}{\partial \theta} \frac{\partial Z}{\partial \theta} 
- 2 X Y Z^2 \frac{\partial^2 Y}{\partial r^2} \right. \\
& + & \left. X Z^2 \left[\frac{\partial Y}{\partial r}\right]^2 
- X Y Z \frac{\partial Y}{\partial r} \frac{\partial Z}{\partial r} 
- 2 X^2 Y Z \frac{\partial^2 Z}{\partial \theta^2} 
+ X^2 Y \left[\frac{\partial Z}{\partial \theta}\right]^2 \right) \\
R_{2 3} & = & 0 \\
R_{3 3} & = & \frac{{\rm e}^{2 \gamma}}{4 X^2 Y^2 Z} \left( 
- 4 X^2 Y Z^2 \frac{\partial^2 \psi}{\partial \theta^2} 
- 2 X Y Z^2 \frac{\partial \psi}{\partial \theta} 
\frac{\partial X}{\partial \theta} 
+ 2 X^2 Z^2 \frac{\partial \psi}{\partial \theta} 
\frac{\partial Y}{\partial \theta} \right. \\
& - & \left. 2 X^2 Y Z \frac{\partial \psi}{\partial \theta} 
\frac{\partial Z}{\partial \theta} 
- 4 X Y^2 Z^2 \frac{\partial^2 \psi}{\partial r^2} 
+ 2 Y^2 Z^2 \frac{\partial \psi}{\partial r} \frac{\partial X}{\partial r} 
- 2 X Y Z^2 \frac{\partial \psi}{\partial r} 
\frac{\partial Y}{\partial r} \right. \\
& - & \left. 2 X Y^2 Z \frac{\partial \psi}{\partial r} 
\frac{\partial Z}{\partial r} 
- X Y Z \frac{\partial X}{\partial \theta} \frac{\partial Z}{\partial \theta} 
+ Y^2 Z \frac{\partial X}{\partial r} \frac{\partial Z}{\partial r} 
+ X^2 Z \frac{\partial Y}{\partial \theta} 
\frac{\partial Z}{\partial \theta} \right. \\
& - & \left. X Y Z \frac{\partial Y}{\partial r} \frac{\partial Z}{\partial r} 
- 2 X^2 Y Z \frac{\partial^2 Z}{\partial \theta^2} 
+ X^2 Y \left[\frac{\partial Z}{\partial \theta}\right]^2 
- 2 X Y^2 Z \frac{\partial^2 Z}{\partial r^2} \right. \\
& + & \left. X Y^2\left[ \frac{\partial Z}{\partial r}\right]^2 \right)
\end{eqnarray*}

\newpage

\noindent
The scalar curvature is given by

\begin{eqnarray*}
R & = & \frac{{\rm e}^{2(\gamma - \psi)}}{2 X^2 Y^2 Z^2} \left(
4 X^2 Y Z^2 \frac{\partial^2 \psi}{\partial \theta^2} 
+ 4 X^2 Y Z^2 \left[\frac{\partial \psi}{\partial \theta}\right]^2 
+ 2 X Y Z^2 \frac{\partial \psi}{\partial \theta} 
\frac{\partial X}{\partial \theta} \right. \\
& - & \left. 2 X^2 Z^2 \frac{\partial \psi}{\partial \theta} 
\frac{\partial Y}{\partial \theta} 
+ 2 X^2 Y Z \frac{\partial \psi}{\partial \theta} 
\frac{\partial Z}{\partial \theta} 
+ 4 X Y^2 Z^2 \frac{\partial^2 \psi}{\partial r^2} 
+ 4 X Y^2 Z^2 \left[\frac{\partial \psi}{\partial r}\right]^2 \right. \\ 
& - & \left. 2 Y^2 Z^2 \frac{\partial \psi}{\partial r} 
\frac{\partial X}{\partial r} 
+ 2 X Y Z^2 \frac{\partial \psi}{\partial r} \frac{\partial Y}{\partial r} 
+ 2 X Y^2 Z \frac{\partial \psi}{\partial r} \frac{\partial Z}{\partial r} 
- 4 X^2 Y Z^2 \frac{\partial^2 \gamma}{\partial \theta^2} \right. \\ 
& - & \left. 2 X Y Z^2 \frac{\partial \gamma}{\partial \theta} 
\frac{\partial X}{\partial \theta} 
+ 2 X^2 Z^2 \frac{\partial \gamma}{\partial \theta} 
\frac{\partial Y}{\partial \theta} 
- 4 X Y^2 Z^2 \frac{\partial^2 \gamma}{\partial r^2} 
+ 2 Y^2 Z^2 \frac{\partial \gamma}{\partial r} 
\frac{\partial X}{\partial r} \right. \\  
& - & \left. 2 X Y Z^2 \frac{\partial \gamma}{\partial r} 
\frac{\partial Y}{\partial r} 
+ 2 X Y Z^2 \frac{\partial^2 X}{\partial \theta^2} 
- Y Z^2 \left[\frac{\partial X}{\partial \theta}\right]^2 
- X Z^2 \frac{\partial X}{\partial \theta} 
\frac{\partial Y}{\partial \theta} \right. \\ 
& + & \left. X Y Z \frac{\partial X}{\partial \theta} 
\frac{\partial Z}{\partial \theta} 
- Y Z^2 \frac{\partial X}{\partial r} \frac{\partial Y}{\partial r} 
- Y^2 Z \frac{\partial X}{\partial r} \frac{\partial Z}{\partial r} 
- X^2 Z \frac{\partial Y}{\partial \theta} 
\frac{\partial Z}{\partial \theta} \right. \\ 
& + & \left. 2 X Y Z^2 \frac{\partial^2 Y}{\partial r^2} 
- X Z^2 \left[\frac{\partial Y}{\partial r}\right]^2 
+ X Y Z \frac{\partial Y}{\partial r} \frac{\partial Z}{\partial r} 
+ 2 X^2 Y Z \frac{\partial^2 Z}{\partial \theta^2} \right. \\ 
& - & \left. X^2 Y \left[\frac{\partial Z}{\partial \theta}\right]^2 
+ 2 X Y^2 Z \frac{\partial^2 Z}{\partial r^2} 
- X Y^2 \left[\frac{\partial Z}{\partial r}\right]^2 \right)
\end{eqnarray*}

\end{document}